# Understanding YTHDF2-mediated mRNA Degradation By m$^6$A-BERT-Deg

Ting-He Zhang[1,2], Sumin Jo[1,3], Michelle Zhang[4], Kai Wang[5,6], Shou-Jiang Gao[1,7] and Yufei Huang[1,2,3*]

[1]Cancer Virology Program, UPMC Hillman Cancer Center, University of Pittsburgh School of Medicine, Pittsburgh, PA, USA

[2]Department of Medicine, University of Pittsburgh School of Medicine, Pittsburgh, PA, USA

[3]Department of Electrical and Computer Engineering, Swanson School of Engineering, University of Pittsburgh, Pittsburgh, PA, USA

[4]Department of Electrical and Computer Engineering, The University of Texas at San Antonio, San Antonio, TX 78249, USA

[5]Raymond G. Perelman Center for Cellular and Molecular Therapeutics, Children's Hospital of Philadelphia, Philadelphia, PA, 19104, USA

[6]Department of Pathology and Laboratory Medicine, Perelman School of Medicine, University of Pennsylvania, Philadelphia, PA, 19104, USA

[7]Department of Microbiology and Molecular Genetics, University of Pittsburgh School of Medicine, Pittsburgh, PA, USA

* Corresponding authors: Yufei Huang, Email: YUH119@pitt.edu

## Abstract

N6-methyladenosine (m$^6$A) is the most abundant mRNA modification within mammalian cells, holding pivotal significance in the regulation of mRNA stability, translation, and splicing. Furthermore, it plays a critical role in the regulation of RNA degradation by primarily recruiting the YTHDF2 reader protein. However, the selective regulation of mRNA decay of the m$^6$A-methylated mRNA through YTHDF2 binding is poorly understood. To improve our understanding, we developed m$^6$A-BERT-Deg, a BERT model adapted for predicting YTHDF2-mediated degradation of m$^6$A-methylated mRNAs. We meticulously assembled a high-quality training dataset by integrating multiple data sources for the HeLa cell line. To overcome the limitation of small training samples, we employed a pre-training-fine-tuning strategy by first performing a self-supervised pre-training of the model on 427,760 unlabeled m$^6$A site sequences. The test results demonstrated the importance of this pre-training strategy in enabling m$^6$A-BERT-Deg to outperform other benchmark models. We further conducted a comprehensive model interpretation and revealed a surprising finding that the presence of co-factors in proximity to m$^6$A sites may disrupt YTHDF2-mediated mRNA degradation, subsequently enhancing mRNA stability. We also extended our analyses to the HEK293 cell line, shedding light on the context-dependent YTHDF2-mediated mRNA degradation.

**1 Introduction**

N6-methyladenosine (m$^6$A) is the most abundant internal modification found in mRNA molecules, and it impacts almost all stages of the mRNA life cycle, including splicing, export, translation, and especially, mRNA stability. By regulating mRNA stability, m$^6$A plays a crucial role in controlling various cellular and physiological processes, including spermatogenesis, embryogenesis, cortical neurogenesis, and cancer stem cells in acute myeloid leukemia. [1, 2].

Recent studies have highlighted the role of the m$^6$A reader protein, YTH N6-Methyladenosine RNA Binding Protein 2 (YTHDF2), in regulating mRNA degradation. YTHDF2 can trigger the deadenylation of m$^6$A-containing RNAs by recruiting the CCR4/NOT deadenylase complex to m$^6$A-containing mRNA [3, 4]. Additionally, YTHDF2 collaborates with heat-responsive protein 12 (HRSP12) to facilitate the interaction between YTHDF2 and RNase P/MRP for rapid degradation of the targeted mRNAs [5]. However, not all mRNAs bound by YTHDF2 undergo degradation. Wang et al. compared the mRNA lifetimes in YTHDF2 knockdown vs. control samples in HeLa cells and found that out of approximately 3.6K genes bound by YTHDF2, and 626 genes exhibited an increased mRNA half-life [6]. This finding highlights the selective nature of YTHDF2-mediated mRNA degradation, indicating that YTHDF2-mediated mRNA degradation may be affected by additional RNA Binding Proteins (RBPs). However, our understanding of the selective degradation mechanism mediated by YTHDF2 is still elusive.

To investigate this further, we propose a computational strategy through predicting YTHDF2-mediated mRNA degradation based on mRNA sequences surrounding an $m^6A$ site bound by YTHDF2. We hypothesized that a model capable of differentiating YTHDF2-bound $m^6A$ sites that regulate mRNA degradation from those that do not impact RNA stability could inform the bindings of RBPs involved in YTHDF2-related mRNA degradation. To build the model, we adapted the highly successful language model Bidirectional Encoder Representations from Transformers (BERT) [20] to mRNA sequences and designed a BERT model tailored for $m^6A$-related mRNA context, named as $m^6A$-BERT. To overcome the challenges with the limited training sequences of $m^6A$ sites associated with mRNA degradation, we pre-trained $m^6A$-BERT on 427,760 $m^6A$-containing mRNA sequences from 24 tissue/cell lines and then fine-tuned the model using a training dataset derived from $m^6A$ sites, YTHDF2 binding and mRNA half-life data. We named this fine-tuned version of $m^6A$-BERT as $m^6A$-BERT-Deg. Notably, $m^6A$-BERT-Deg demonstrated improved performance over other state-of-the-art deep learning and machine learning models on YTHDF2-mediated mRNA degradation prediction. We subsequently devised a model interpretation scheme for $m^6A$-BERT-Deg that revealed a potential mechanism of the selective regulation of mRNA stability through the disruption of YTHDF2-mediated mRNA degradation. We further applied $m^6A$-BERT-Deg to predict condition-specific regulation of YTHDF2-mediated mRNA degradation in HEK293T cells and obtained prediction results that were supported by *m6A-express*. Additionally, disruption of YTHDF2-mediated mRNA degradation was also observed in HEK293T cells.

## 2 Materials and methods

### 2.1 Data

#### 2.1.1 Dataset for m$^6$A-BERT pre-training

427,760 human m$^6$A sites were collected from m$^6$A-AtlasV2 [7-9]. These sites were identified by single-base m$^6$A profiling technologies including m6A-CLIP-seq, miCLIP, m6A-SAC-seq, and others. To prepare the dataset for model pre-training, we mapped these m$^6$A sites from the genome to their corresponding transcripts. We then extended the regions by 250 base pairs (bp) on each side of the site and extracted 501bp site sequences. We obtained 427,760 m$^6$A-containing sequences. These sequences were used for m$^6$A-BERT pre-training.

#### 2.1.2 Dataset for fine-tuning m$^6$A-BERT-Deg

We assembled a dataset that includes so-called positive and negative sequences, both of which are m$^6$A site sequences bound by YTHDF2. The positive sequences are those whose mRNAs show an increased half-life in YTHDF2 depleting *vs.* wild-type (WT) cells and the negative sequences are those whose mRNAs do not.

To collect m$^6$A site sequences bound by YTHDF2, we used YTHDF2 PAR-CLIP data from HeLa cells (GSE49339) [6], which included 52,823 binding sites of YTHDF2 in 3,611 genes identified from three biological replicates. We next extracted 61,369 m$^6$A sites in HeLa cells from m$^6$A-AtlasV2 and identified 8,402 sites that overlapped with YTHDF2 binding sites.

These sites were used to construct our training dataset. Specifically, to determine the positive sites or sites that regulate mRNA decay, we obtained the data on mRNA half-life in YTHDF2 knockdown (siYTHDF2) and control (siControl) HeLa cells (GSE49339) [6]. We next selected 626 genes with the averaged log fold change ($log_2(siYTHDF2/siControl)$) greater than 1 as candidate genes for identifying positive sites because they are associated with an increased lifetime and therefore could regulate mRNA decay. Out of the 8,402 m$^6$A sites that have YTHDF2 binding, 666 of them were in CDS and 3'UTR regions of these 626 genes with increased half-life and therefore treated as positive training sites. Note that sites in the 5'UTR of these genes were filtered out because YTHDF2-binding at the 5' UTR has been reported to facilitate protein translation rather than mRNA degradation [10]. The remaining 7,726 m$^6$A sites that were not in these 626 genes were considered negative sites. We further extracted 501bp sequences centered at positive/negative sites. The sequences were trimmed if they exceeded the transcript length. To prevent having similar sequences in both training and testing sets, we applied CD-HIT [11] to remove similar sequences within both the positive and negative sets, using a cutoff threshold of 90%. Finally, we obtained 485 positive sequences and 5,602 negative sequences. To create a balanced training set and test set, we randomly selected 485 sequences from the negative set as the final set of negative sequences.

All the m$^6$A sites were based on the hg38 genome assembly, while the YTHDF2 PAR-CLIP data, originally based on the hg19 assembly, were converted to the hg38 assembly using the liftOver tool [12]. Gene annotations were obtained from GENCODE v36 [13]. All sequences were denoted by DNA-encoded nucleotides, i.e., "ATGC" instead of "AUGC".

## 2.2 The m⁶A-BERT model

Inspired by DNABERT [14], m⁶A-BERT was modified from BERT and followed a two-step framework consisting of pre-training and fine-tuning. During the pre-training, m⁶A-BERT initialized the model weights using the large collection of pre-training m⁶A-site sequences, thereby allowing it to capture the specific context and characteristics of m⁶A modifications. Afterward, in the fine-tuning phase, the pre-trained m⁶A-BERT was further trained for predicting mRNA degradation using the training dataset with limited positive and negative sequences.

m⁶A-BERT includes a token embedding layer, a position embedding layer, and 12 transformer blocks followed by a classification layer. All pre-training and training sequences were first tokenized into a sequence of $K$-mers and passed into the embedding layer before being fed to the 12 transformer blocks (Fig. 1). We describe the details of each step next.

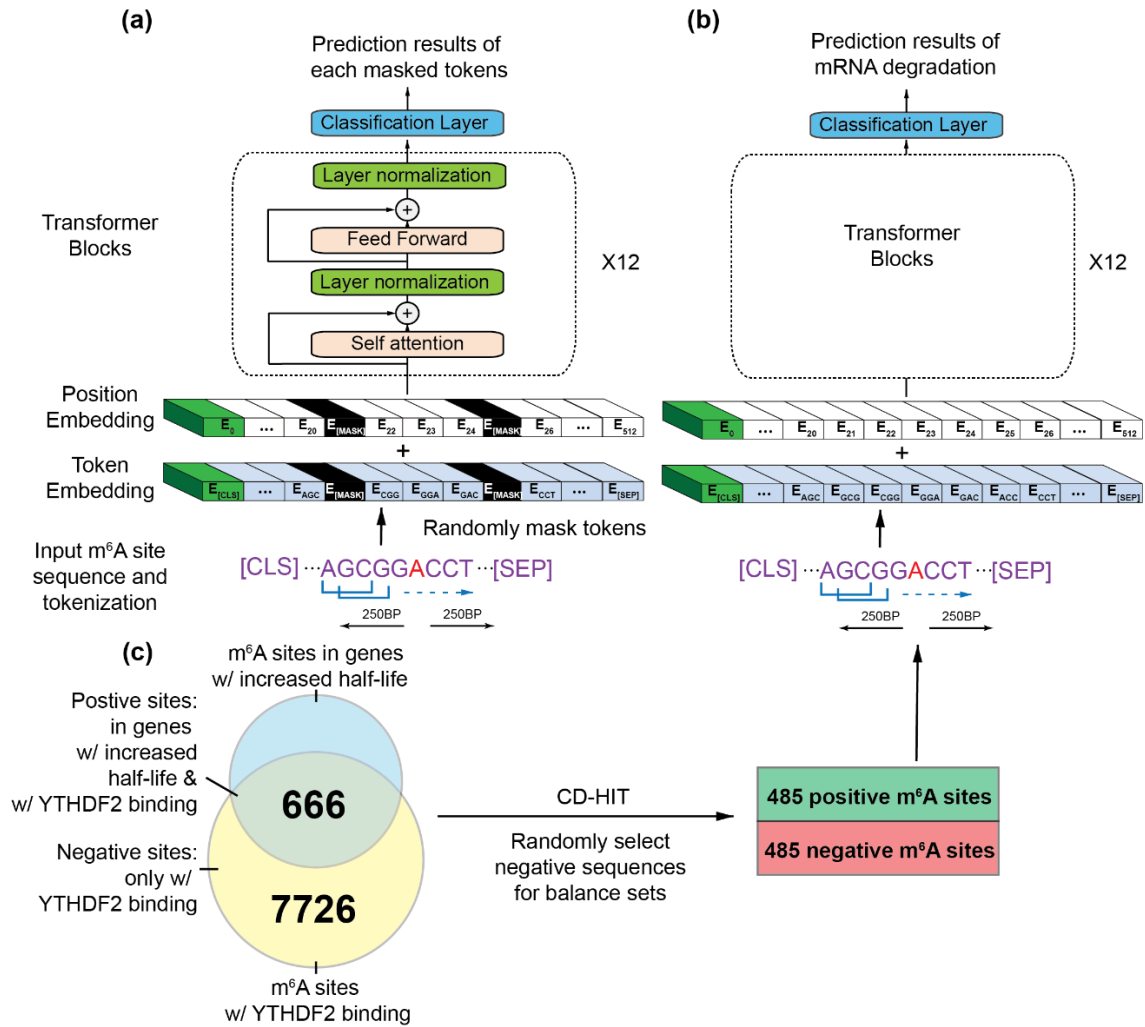

Figure 1. Illustration of the model architecture and the training process for m⁶A-BERT-Deg (*K* =3). (a) m⁶A-BERT was pre-trained on human single-based m⁶A sites using the random masking; (b) m⁶A-BERT was fine-tuned on the mRNA degradation dataset to establish m⁶A-BERT-Deg; (c) The preparation of the training dataset for m⁶A-BERT-Deg. m⁶A sites with YTHDF2 binding were divided into those that may or may not regulate mRNA decay, resulting in 485 positive and negative m⁶A sites, respectively. Each site was extended 250 bp on both sides to obtain a 501 bp sequence, which is the input sequence of m⁶A-BERT-Deg.

### 2.2.1 Tokenization

The goal of tokenization is to represent each input RNA sequence as a sequence of words or tokens as required by the BERT model. We denoted a token as a *K*-mer sequence and tokenized the input sequence by using a sliding window of length *K* [15]. For example, if we have an input sequence 'AGCGGA', this tokenization can generate four 3-mers: 'AGC', 'GCG', 'CGG', 'GGA' or three 4-mers: 'AGCG', 'GCGG', 'CGGA'. Each *K*-mer was then treated as a word or token of the sequence, similar to that in DNABERT. Special tokens were added to each input sequence including [CLS] at the beginning, which represents the classification token, [SEP] at the end, which indicates the separation token, and [MASK] for masked tokens. Since our input sequence had a length less than 501, [PAD] tokens were also added before the [SEP] token to pad the sequence to a length of 501, maintaining consistency with the input size required by the BERT model.

Since different values of *K* can result in different tokenization of sequences, we pre-trained four versions of m$^6$A-BERT for each *K*=3,4,5,6. These models were pre-trained to capture the varying context and features of the m$^6$A site sequences at different granularities.

### 2.2.2 Pre-training

To capture the patterns in m$^6$A site sequences, we employed the pre-training pipeline from DNABERT and utilized the Masked Language Model (MLM) task to pre-train m$^6$A-BERT. During pre-training, 15% of tokens (except special tokens) in the input sequences were randomly masked and the model was trained to predict these masked tokens based on the remaining tokens. Initially, the MLM rate was set at 15% and the model was trained until convergence, which typically occurred around 100,000 to 120,000 steps. Subsequently, we increased the MLM rate to 20% and continued training for an additional 20,000 steps. To mitigate the risk of overfitting during the pre-training stage, we randomly selected 90% of the pre-training dataset for training and the remaining 10% for validation.

Similar to BERT [16], our model structure consists of 12 transformer layers containing 768 hidden units and 12 attention heads (Fig 1). The parameter settings remained consistent across all the pre-trained models. Each model was pre-trained on 2 Tesla A100 GPUs.

### 2.2.3 Fine-tuning

To adapt the pre-trained m$^6$A-BERT for predicting mRNA degradation, we employed a binary classification layer on top of the pre-trained model and fine-tuned it using the training dataset. The output of the classification layer is "1" if the input site sequence is predicted to regulate mRNA degradation" and '0', otherwise.

To fine-tune the model, we utilized a data splitting strategy, where 70% of the data was designated as the training set, 20% as the testing set, and the remaining 10% as the validation set. We fine-tuned the model with 100 epochs and used the model with the highest AUC on the validation set for further analysis. The learning rate for fine-tuning was 1e-5.

Since m$^6$A-BERT with K = 3, 4, 5, 6 exhibited closely similar performances with minor fluctuations, we consistently reported results of *K*-mer = 3 in all our experiments, as it consistently yielded the best performance.

### 2.3 Performance metrics

To comprehensively evaluate the performance of our model, we chose 5 metrics including ACC (accuracy), MCC (Matthews correlation coefficient), AUC (the area under the ROC curve), precession, and recall. MCC is a reliable metric that produces a higher score only if the model performs more accurate predictions. ACC, MCC, and precision are defined as

$$ACC = \frac{TP + TN}{TP + TN + FP + FN} \qquad (1)$$

$$MCC = \frac{TP \cdot TN - FP \cdot FN}{\sqrt{(TP + FP) \cdot (TP + FN) \cdot (TN + FP) \cdot (TN + FN)}} \qquad (2)$$

$$Precision = \frac{TP}{TP + FP} \qquad (3)$$

where TP, TN, FP, and FN are the numbers of true positives, true negatives, false positives, and false negatives, respectively. The ROC curve is the curve of FPR (false positive rate) vs. TPR (true positive rate, also known as recall) with

$$FPR = \frac{FP}{TN + FP} \qquad (4)$$

and

$$TPR = Recall = \frac{TP}{TP+FN}. \qquad (5)$$

AUC was computed as the area under the ROC curve.

## 2.4 Model interpretation

We deployed model interpretation approaches to understand how m⁶A-BERT makes a prediction and uncover key sequence motifs that might inform the binding of co-factor RBPs.

### 2.4.1 Attribution scores

Attribution scores describe the contribution of input tokens or attention weights to the prediction [17]. Given the model with an input token $x$ and a reference token $x_{ref}$, an attribution score, integral gradient (IG), is computed as

$$IG(x) = (x - x_{ref}) \times \int \frac{\partial F\left(x + \alpha \times (x - x_{ref})\right)}{\partial x} d\alpha \qquad (6)$$

where the integral is taken along a straight line from $x_{ref}$ to $x$ parameterized by the parameter α. The reference input was created as a sequence of [PAD] tokens with [CLS] and [SEP] tokens at the beginning and end of sequences. We computed the attribution score on the embedding layer by using Captum [18]. To obtain the attribution score for each nucleotide, we calculated the score by averaging the sum of attribution scores from all related tokens based on their respective counts.

### 2.4.2 Hypergeometric test for motif discovery

To identify biologically significant motifs within a given set of sequences (positive or negative), we adjusted and applied the motif analysis tool from DNABERT. Initially, we extracted contiguous nucleotide regions exhibiting consistently higher attribution scores than the average for both positive and negative sequences. These high-scoring nucleotide regions were then considered as preliminary motif candidates. Notably, multiple motif candidates could be extracted from a single sequence.

Next let random variable $X$ be the frequency of motif candidates with same nucleotide appearing within the given set. We assume $X$ followed hypergeometric distribution [19] $X \sim Hypergeometric(N, K, n)$:

$$P(X = k) = \frac{\binom{K}{k}\binom{N-k}{n-k}}{\binom{N}{n}} \tag{7}$$

where $N$ is the total number of motif candidates, $K$ is the number of motif candidates in the given set of sequences, and $n$ is the frequency of motif candidates with the same nucleotide appearing in all sequences. The null hypothesis $H_0$ for this test: there is no significant difference between the frequency of motif candidates with the same nucleotide within the given set of sequences and all sequences. Motif candidates with an adjusted $p$-value $< 0.05$ were selected as significantly distinguishing motifs. The motifs candidates with a frequency of less than 5 were excluded from consideration.

Once the distinguishing motifs were identified, we extracted the 24 bp long sequences with these motifs at the center from each sample. These 24 bp long sequences were used as input for TOMTOM[44] to match against known motifs.

## 3 Results

### 3.1 Pre-training enables m$^6$A-BERT-Deg to outperform other models

We systematically assessed the performance of m$^6$A-BERT-Deg and compared its performance with several baseline models. To demonstrate the effectiveness of pre-training, we trained BERT-baseline, a BERT model trained directly using the training dataset without pre-training. We also trained DNABERT-Deg, a fine-tuned DNABERT, which was pre-trained using DNA sequences. Comparing m$^6$A-BERT-Deg against DNABERT-Deg can reveal the impact of using m$^6$A site sequences over DNA sequences for pre-training. Additionally, we trained iDeepMV-Deg [20] and CNN+LSTM-Deg [21], two existing sequence-based deep learning models using the training dataset and used their performance to evaluate the improvement gained by the BERT model. To provide a fair comparison, we selected the best-performing feature for iDeepMV-Deg, which in this case was the dipeptide component of RNA and used its performance for comparison. For the BERT-baseline, DNABERT-Deg, and our m$^6$A-BERT-Deg models, we selected the model with the best performance among the four *K*-mer models for each respective method. For all the models, we performed 5-fold cross-validation (CV) to compute their prediction performance (Table 1).

Table 1. Performance comparison of m$^6$A-BERT-Deg and other benchmark models in 5-fold CV

|  | ACC | AUC | MCC | Precision | Recall |
|---|---|---|---|---|---|
| CNN+LSTM-Deg | 61.75%±5.71% | 71.11%±1.26% | 0.2697±0.0554 | 66.78%±6.98% | 56.25%±21.57% |
| iDeepMV-Deg | 67.73%±1.65% | 70.12%±4.37% | 0.3683±0.0379 | 66.00%±4.73% | 76.58%±11.33% |
| BERT-baseline | 67.31%±1.06% | 72.76%±1.33% | 0.3492±0.0185 | 67.59%±0.80% | 67.31%±1.06% |
| DNABERT-Deg | 68.92%±1.93% | 75.98%±2.06% | 0.3838±0.0381 | 69.44%±1.93% | 68.92%±1.93% |
| m$^6$A-BERT-Deg | **71.65%±2.06%** | **77.47%±1.98%** | **0.4335±0.0411** | **71.70%±2.04%** | **71.65%±2.06%** |

Overall, m$^6$A-BERT-Deg produces the best performance among all tested models with all metrics considered. Interestingly, BERT-baseline does not show a clear advantage over iDeepMV-Deg. The small training samples probably were insufficient to fully leverage the capabilities of the BERT model. In contrast, m$^6$A-BERT-Deg shows a 4% gain in ACC and AUC over the BERT-baseline, clearly demonstrating the benefit of BERT-based pre-training. Furthermore, m$^6$A-BERT-Deg outperforms DNABERT-Deg, highlighting the advantage of pre-training with m$^6$A site sequences. This focus on m$^6$A site sequences as opposed to DNA sequences allows m$^6$A-BERT-Deg to capture more nuanced, m$^6$A-specific context and semantic features. Such unique features might encompass critical information like RBP binding sites, which could prove essential for accurately predicting mRNA degradation. Taken together, these findings revealed the superior performance of the proposed m$^6$A-BERT-Deg, underlining the significance of pre-training using m$^6$A site sequences.

## 3.2 Model interpretation revealed enriched RBP bindings in negative m$^6$A sequences

We conducted model interpretation of m$^6$A-BERT-Deg to investigate discriminative features that could inform YTHDF2-associated co-factors that mediate mRNA degradation. We first computed the attribution scores of m$^6$A-BERT-Deg's embedding layer by using IG. These attribution scores can inform the contribution of each input token to the prediction, where tokens with higher attribution scores would suggest a greater impact on the prediction. To visualize the distribution of these attribution scores, we generated a heatmap to display the scores' distribution across all positive and negative samples. This heatmap provides a visual representation of the contribution of each token in the prediction process, facilitating the identification of m⁶A-BERT-Deg's discriminative features (Fig 2).

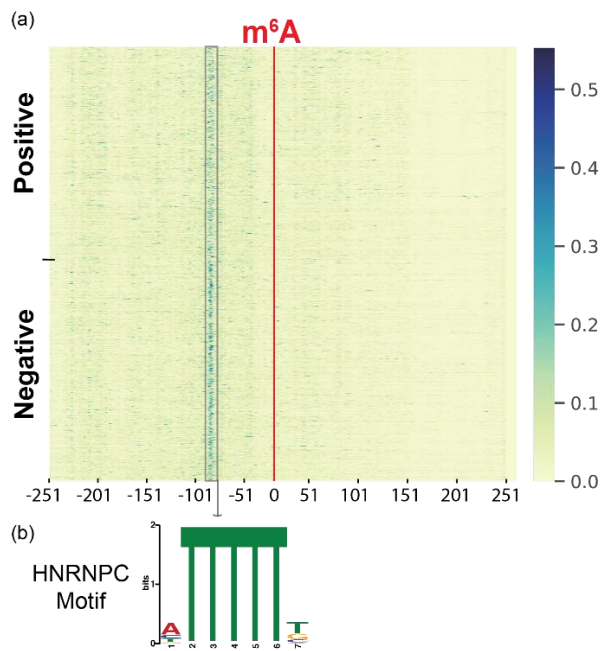

Figure 2. (a) Heatmap of attribution scores for all samples. The regions with high attribution scores are selected as motif candidates and used for RBPs enrichment analysis in positive and negative sets; (b) HNRNPC exhibits a substantial enrichment in regions with both high and the highest attribution scores (80bp upstream of m⁶A) in negative sets. There were notable number of sequences within the negative dataset that corresponded to the HNRNPC motif.

Careful examination of this heatmap revealed that the m⁶A sites, located at the center ($0^{th}$ bp) of the sequence (Fig. 2), generally were associated with low attribution scores. This was expected because both positive and negative training sequences contain m⁶A sites. However, regions upstream of the m⁶A sites show higher attribution scores, indicating that these upstream regions might play a more significant role in influencing mRNA degradation (Fig. 2a). Notably, the region around 80bp upstream (-80 bp in Fig 2a) of m⁶A exhibited the highest attribution scores, suggesting sequences within this region could contain discriminate features, likely associated with binding motifs of YTHDF2 co-factors. To systematically identify these co-factors, we first determined regions of contiguous nucleotides with consistently higher attribution scores than the average. Then, we conducted a direct mapping of RBP binding in the HeLa cell line from POSTAR3 to these regions with higher attribution scores in positive and negative sequences [22] and obtained 23 RBPs (Supplementary Table S1). We then employed the Fisher exact test to determine RBPs that have significantly enriched binding frequencies in positive and negative site sequences. While EIF4A3 was the only RBP determined to have significantly higher binding frequencies in positive sequences, ten RBPs, namely HuR, HNRNPC, HNRNPU, U2AF2, TIA1, CSTF2, HNRNPA1, ALYREF, TIAL1, and U2AF65 showed significantly higher frequencies in negative sequences (adjusted $p$-value < 0.05). These overwhelmingly more enriched bindings of RBPs in negative sequences not associated with mRNA decay could suggest a possibility that their binding in the neighborhood of m$^6$A sites might disrupt the role of YTHDF2 in promoting mRNA degradation. Indeed, HuR is a well-known RNA stability factor that stabilizes mRNA levels by binding to target mRNAs[23]. HuR is a known m$^6$A reader[24], which has been shown to be blocked from binding to m$^6$A sites to facilitate m$^6$A-mediated decay [25]. We have also shown previously that HuR promoted the stability of its target mRNAs in an m$^6$A-dependent manner and HuR-binding sites were in close proximity to m$^6$A-binding motifs (RRACH) [26]. Besides HuR, HNRNPC has also been demonstrated to stabilize mRNA by interacting with poly-U tracts in the 3'-UTR or 5'-UTR of mRNA [27, 28]. HNRNPC is a member of the hnRNPs family and is recognized as an important m$^6$A reader protein. HNRNPC is known for its function in shaping local structure in m$^6$A-containing mRNA [29, 30]. Another hnRNPs family protein, HNRNPU, was also an enriched RBP in negative sequences. HNRNPU is a multifunctional RNA- and DNA-binding protein with a central role in regulating pre-mRNA splicing, mRNA stability, and translation. [31]. It can enhance gene expression by stabilizing mRNA [32]. It is particularly involved in processes such as transcription and the control of mRNA stability.

HNRNPA1, also a member of the hnRNP family, plays a pivotal role in regulating diverse RNA processing mechanisms, including splicing, stability and etc. [33] An additional facet of its functionality involves the enhancement of mRNA stability through targeted binding to specific 3'-UTR sites [34]. BioGRID shows that HNRNPA1 exhibits an interaction with YTHDF2, potentially forming a complex or mutually binding to one another [35]. This interaction could impede the mRNA decay process medicated by YTHDF2. Other RBPs also demonstrated substantial connections with mRNA stability. For instance, ALYREF, a heat-stable nuclear chaperone, enhances mRNA stability by binding to specific regions of mRNA[36-38]. U2AF2 is a non-snRNP protein and it facilitates the binding of U2 snRNP to the pre-mRNA branch site. It shows the capability to associate with distinct internal RNA elements, thereby enhancing the mRNA stability [39, 40]. TIAL1, known as TIA1 Cytotoxic Granule Associated RNA Binding Protein Like 1, plays multifaceted roles, including its ability to enhance mRNA stability [41, 42]. Taken together, our results revealed much higher frequencies of RBP bindings in negative $m^6A$ sequences than in positive sequences. Many of these enriched RBPs showed a functional propensity to promote RNA stability, implying a potential role they play in disrupting YTHDF2-mediated mRNA degradation.

To further confirm this finding, we applied the hypergeometric test to identify motifs enriched in negative *vs.* positive training sequences (adjusted *p*-value < 0.05). We identified 4 significant motifs through matching them with 102 known human RBPs' motifs in Ray2013 [43] using TOMTOM [44], two were found to match the motifs of HuR (adjusted *p*-value = 0.0081) and HNRNPC (adjusted *p*-value= 0.0513) (Supplementary Figure S1). This result is consistent with the finding above as these two RBPs were among the enriched RBPs in negative $m^6A$ sequences above and are $m^6A$ readers that have been shown to enhance mRNA stability.

**3.3 Predicting YTHDF2-mediated mRNA degradation in HEK293T cells using m$^6$A-BERT-Deg**

We next applied m$^6$A-BERT-Deg to predict context-specific YTHDF2-mediated mRNA degradation in HEK293T cells. To this end, we collected 120,460 single base m$^6$A sites of HEK293T cells from m$^6$A-AtlasV2 and 118,233 YTHDF2 binding regions from the PAR-CLIP dataset (GSE78030) [22, 45] in HEK293T cells. To generate input sequences specific to the HEK293T cell line, we selected m$^6$A sites that were inside YTHDF2 binding sites and then extracted 501bp sequences centered at each site. This process generated 5,047 candidate m$^6$A site sequences. We then applied m$^6$A-BERT-Deg to these sequences and m$^6$A-BERT-Deg predicted 1,281 of them to potentially regulate mRNA degradation of their methylated genes (Supplementary Table S3). As direct evaluation of this prediction was hampered by the lack mRNA half-life for the HEK293T cell line, we resorted to an alternative approach by using *m$^6$A-express*[46]. *m$^6$A-express* is a Bayesian hierarchical model that predicts if a site regulates gene expression by modeling coordinated changes between m$^6$A methylation levels and the gene expressions due to this regulation using MeRIP-seq data. In this case, we applied *m$^6$A-express* to MeRIP-seq samples (GSE182607 [47]) of METTL3 knockout (KO) in HEK293T cells. An m$^6$A site that induces mRNA degradation in HEK293T cells likely exhibits changes in the expression of its methylated mRNAs in either condition and would be predicted by *m$^6$A-express*. To assess m$^6$A-BERT-Deg's prediction, we first examined the MeRIP-seq profiles of 5,047 candidate m$^6$A sites and retained, for the METTL3-KO condition, 236 positive and 479 negative sites whose corresponding genes had a peak with sufficient reads (>10 reads for IP samples) in MeRIP-seq samples for both WT and METTL3-KO condition. Then, we applied *m$^6$A-express* and predicted

58 and 57 genes whose gene expression might be regulated by positive and negative m6A sites, respectively (adjusted *p*-value < 0.05). Notably, we found that genes from the positive set exhibited higher enrichment compared to the negative set (Fisher exact test, *p*-value=3.23×10$^{-5}$), indicating positive sets were more enriched with sites that predicted to regulate gene expression. We further examined $β_1$, a key parameter of the *m6A-express* output that would inform the mode (positive or negative) and degree of the impact that m6A methylation have on gene expression. Comparison between $β_1$ values between 58 and 57 genes showed that genes from the positive set exhibited significantly more negative $β_1$ values (one-sided Wilcoxon rank test; adjusted *p*-value = 0.005593) (Fig 3a), suggesting that these positive sites that predicted to regulate gene expression by *m6A-express* likely induce a more pronounced downregulation on gene expression than the negative sites, thus having a higher chance to regulate mRNA degradation than negative sites. Together, these results served to indirectly support the validity of the predicted m6A sequences by m6A-BERT-Deg to potentially regulate mRNA degradation.

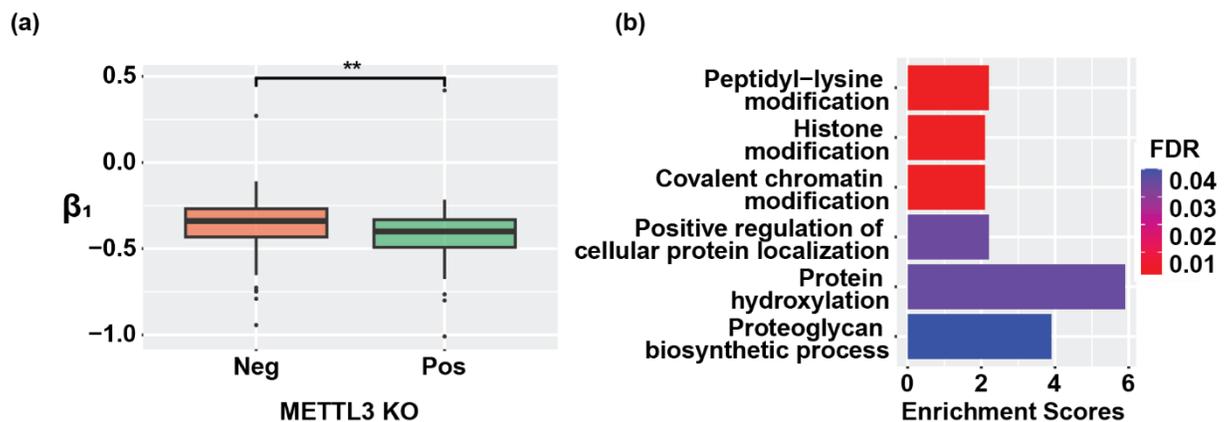

Figure 3. (a) Boxplot for the regulation degree of the methylation level on gene expression for positive sets and negative sets for METTLE3 KO; (b) Gene Ontology (GO) Enrichment analysis for predicted genes that had degradation sites in HEK293T cells.

To understand the functions of these 236 predicted sites, we performed Gene Ontology (GO) enrichment of their methylated genes and revealed a significant enrichment of six key biological functions (adjusted *p*-value < 0.05; Fig 3b) including peptidyl-lysine modification, histone modification, covalent chromatin modification, and protein hydroxylation (Fig. 3b), suggesting that m$^6$A-mediated mRNA degradation in HEK293T cells may be engaged in regulating post-translational modification.

To further understand potential YTHDF2 co-factors involved in regulating mRNA degradation associated with these predicted sites, we conducted the same enrichment of RBP binding sites. We chose only RBPs of HEK293T cells from POSTAR3 and were able to map the sites of 28 RBPs to regions with higher attribution scores in both positive and negative sequences (Supplementary Table S2). We then identified RBPs with significantly enriched binding frequencies in positive and negative site sequences (Fisher exact test, adjusted p-value< 0.05). Like our findings in HeLa cells, we observed a much higher number of RBPs exhibiting enriched binding in negative sequences compared to positive sequences. While ten RBPs— HNRNPC, U2AF2, HNRNPA1, FUS, HNRNPF, RBM10, DDX3X, SRSF1, CHTOP, and ATXN2— demonstrated higher binding frequencies in negative sequences, only two RBPs, HNRNPH and EIF3D, displayed higher binding frequencies in positive sequences. This result further supports the potential role of YTHDF2 co-factors in disrupting RNA decay. Out of the ten RBPs enriched in negative sequences, HNRNPC, U2AF2, and HNRNPA1 were also enriched in HeLa cells. In addition, among RBPs unique to HEK293 cells, RBM10 is known to stabilize various genes including apoptosis related genes [48, 49]. FUS, a versatile RNA-binding protein, has been reported to exert a direct influence on RNA metabolism through promoting mRNA stabilization [50-52]. Taken together, our analysis in HEK293 cells confirmed the potential role of YTHDF2 co-factors in disrupting mRNA decay. It further unveiled the context-dependent nature of these cofactors to interfere with mRNA degradation.

**4 Discussion and conclusion**

We have trained m$^6$A-BERT-Deg to predict YTHDF2-mediated mRNA degradation. To facilitate this training, we assembled a high-quality training dataset comprising 485 positive and 485 negative m$^6$A site sequences for the HeLa cell line. This dataset was created by integrating various data sources including published single-base m$^6$A sites, YTHDF2 binding PAR-CLIP data, and mRNA half-life data. To address the challenge posed by the relatively small training samples, we employed a pre-training strategy, where m$^6$A-BERT underwent self-supervised training using 427,760 m$^6$A site sequences that might or might not be associated with regulating mRNA degradation. Our test results clearly demonstrated the importance of this pre-training strategy. It played a critical role in enabling m$^6$A-BERT-Deg to outperform other state-of-the-art models, resulting in its best performance.

To gain a deeper understanding of the selective YTHDF2-mediated mRNA decay, we conducted model interpretation to identify crucial regions within the training sequences. Surprisingly, when we mapped known RBP binding sites within these regions, we observed a notable enrichment of RBP bindings in negative sequences, in contrast to positive sequences. Remarkably, several of these RBPs, including HuR and HNRNPC have previously been documented to promote mRNA stability in a m$^6$A-dependent manner. These findings suggest that the presence of these RBPs in proximity to m$^6$A sites may disrupt YTHDF2-mediated mRNA degradation, subsequently enhancing mRNA stability. This discovery unveils a potential mechanism that underlies the selective nature of YTHDF2-mediated mRNA degradation.

To demonstrate the utility of m$^6$A-BERT-Deg in predicting context-specific YTHDF2-mediated mRNA degradation, we conducted predictions within the HEK293T cell line. We validated our predictions by leveraging m$^6$A-express analyses from MeRIP-seq samples obtained under wild-type and METTL3 KO conditions. Our analysis suggested that the predicted positive sites were more likely to regulate mRNA degradation than their negative counterparts. We further extended our analyses of YTHDF2 cofactor to HEK293T cells, confirming the discovery that RBP bindings could disrupt YTHDF2-mediated mRNA degradation.

**Acknowledgments**

This research was supported by grants from the National Cancer Institute Informatics Technology for Cancer Research (ITCR) (U01CA279618 to Y. Huang), National Institute of Health (1SC3GM136594-02 to J. Zhang) and National Cancer Institute (CA124332 to S.-J. Gao).

**Code availability**

The detailed instructions and pre-trained models are available at https://github.com/TingheZhang/m6ABERT.